\documentclass[english,twocolumn,floats,showpacs,amsmath,amssymb,prb]{revtex4}
\usepackage[latin9]{inputenc}
\usepackage{graphicx}
\usepackage{amssymb}

\makeatletter

\providecommand{\tabularnewline}{\\}

\usepackage{epsfig}

\usepackage{babel}
\makeatother

\begin{document}

\title{Quantum path-integral study of the phase diagram and isotope effects
of neon }

\author{R. Ram\'{\i}rez }

\author{C. P. Herrero}

\affiliation{Instituto de Ciencia de Materiales de Madrid (ICMM), Consejo Superior
de Investigaciones Cient\'{\i}ficas (CSIC), Campus de Cantoblanco,
28049 Madrid, Spain }

\begin{abstract}
The phase diagram of natural neon has been calculated for temperatures
in the range 17-50 K and pressures between $10^{-2}$ and $2\times10^{3}$
bar. The phase coexistence between solid, liquid, and gas phases has
been determined by the calculation of the separate free energy of
each phase as a function of temperature. Thus, for a given pressure,
the coexistence temperature was obtained by the condition of equal
free energy of coexisting phases. The free energy was calculated by
using non-equilibrium techniques such as adiabatic switching and reversible
scaling. The phase diagram obtained by classical Monte Carlo simulations
has been compared to that obtained by quantum path-integral simulations.
Quantum effects related to the finite mass of neon cause that coexistence
lines are shifted towards lower temperatures when compared to the
classical limit. The shift found in the triple point amounts to 1.5
K, i.e., about 6\% of the triple-point temperature. The triple-point
isotope effect has been determined for $^{20}$Ne, $^{21}$Ne, $^{22}$Ne,
and natural neon. The simulation data show satisfactory agreement
to previous experimental results, that report a shift of about 0.15
K between triple-point temperatures of $^{20}$Ne and $^{22}$Ne.
The vapor-pressure isotope effect has been calculated for both solid
and liquid phases at triple-point conditions. The quantum simulations
predict that this isotope effect is larger in the solid than in the
liquid phase, and the calculated values show nearly quantitative agreement
to available experimental data.
\end{abstract}


\maketitle

\section{Introduction\label{sec:intro}}

The electronic structure of atoms and molecules, encountered either
in gas or condensed phases, is, within the framework established by
the Born-Oppenheimer (BO) approximation, independent of the nuclear
mass. Thus the BO potential energy surface (PES)\emph{, }that represents
the energy landscape for the motion of the nuclei,\emph{ }results
to be\emph{ independent} of the isotope mass. Many processes and properties
depend on the movement of the atoms on the PES. Then, isotope effects
may exist because the \emph{kinetic energy} of the nuclei is mass
dependent according to the laws of quantum mechanics. Thus, although
potential energy parameters are isotope independent, the kinetic energy
parameters are not.\citep{jancso74} Phase coexistence is an equilibrium
property that depends on the kinetic energy of the atoms (think of
a liquid in equilibrium with its vapor). Hence, the phase diagram
of a given substance should depend on its nuclear mass and then display
an isotope effect. We stress that this behavior is strictly of quantum
mechanical nature, i.e., the classical limit in statistical thermodynamics
\emph{incorrectly} predicts that phase coexistence is a mass independent
property. This erroneous prediction can be considered as a consequence
of the \emph{equipartition theorem} for the kinetic energy, that establishes
that in the classical limit this energy is only a function of temperature
and is therefore independent of the atomic mass.

The properties of rare-gas solids or fluids usually change as a function
of the atomic mass in a systematic fashion, that has been used to
study effects of the zero-point energy and isotope mass. There have
been many studies of quantum effects in the phase coexistence of neon,
which were initially motivated by the observation that neon deviates
from the principle of corresponding states, that is however rather
accurately followed by argon, krypton, and xenon.\citep{guggenheim45}
The calculation of \emph{quantum corrections} to the classical limit
of the phase diagram of neon is thus interesting because allows us
to quantify systematic errors of treating the neon atoms as particles
moving classically in the PES. However, we stress that these quantum
corrections  do not represent any kind of measurable property, i.e.,
they can not be directly compared to experimental data, because there
is no way to perform measurements of the phase behavior in the classical
limit. In this respect, the calculation of isotope effects in the
phase diagram of neon results specially interesting because they can
be directly compared to experimental data, offering then a direct
test of the capability of the theoretical approach. Neon has three
stable isotopes ($^{20}$Ne, $^{21}$Ne, and $^{22}$Ne) and the experimental
study of isotope effects in its phase diagram is  historically interesting.
The observation by Keesom and Dijk\citep{keesom23} in 1931 that $^{20}$Ne
was more volatile than $^{22}$Ne at temperatures just above the triple
point was the first demonstration of an isotope effect on the physical
or chemical properties of matter, and this experiment was considered
to be a proof of the existence of the zero-point energy.\citep{jancso74}
An accurate determination of the vapor pressure difference between
natural neon and pure $^{20}$Ne and $^{22}$Ne isotopes has been
made by Bigeleisen and Roth\citep{bigeleisen68} and Furokawa.\citep{furokawa72}
The vapor pressure isotope effect (VPIE) amounts to about 5\% of the
equilibrium pressure, being slightly larger for solid than for liquid
phases. The triple-point isotope effect (TPIE) has been also measured
for the $^{20}$Ne and $^{22}$Ne isotopes,\citep{clusius60} and
amounts to a temperature shift of 0.15 K. This difference might seem
small, but it is relevant for thermometry as the triple point of natural
neon ($T_{p}=24.553$ K) is used as fixed-point temperature standard
in the International Temperature Scale of 1990 (ITS-90), and any deviation
of the isotopic composition of natural neon leads to ambiguity in
the reference temperature.\citep{pavese08} 

The most employed quantum approaches to the phase diagram of neon
are based either on perturbational methods using the Wigner-Kirkwood
expansion of the quantum partition function in powers of $\hbar$,
or on computer simulations using the path integral (PI) formulation
of statistical mechanics. We are not aware of PI simulations of isotopic
effects in the phase coexistence of neon. However, PI MC simulations
of isotope effects in the lattice parameter and heat capacity of solid
neon have been reported, \citep{herrero01} showing good agreement
to experimental data. The VPIE of liquid neon has been recently 
studied\citep{lopes03}
 using a classical integral equation theory in order to obtain the
parameters required for the statistical theory of isotope effects
developed by Bigeleisen in 1961.\citep{bigeleisen61} The isotopic
shift in the helium melting curve has been studied by PI MC simulations
at temperatures where nuclear exchange effects associated to the nuclear
quantum statistics are negligible.\citep{barrat89,boninsegni94} Isotope
effects in the solid-liquid coexistence of Lennard-Jones (LJ) solids
have been studied by the method of generating Landau free energy curves.
This method was applied to study the melting curves of solid H$_{2}$
and D$_{2}$, for which quantum effects are expected to be much larger
than for neon, due to its lighter molecular mass. At about 12 MPa
the calculated melting temperature of \emph{ortho}-D$_{2}$ and 
\emph{para}-H$_{2}$
was of 22.3 and 18 K, respectively.\citep{chakravarty00}

In a summary of  theoretical calculations of quantum corrections in
neon, we should mention the study by Hansen and Weis\citep{hansen69}
of the coexistence curve of neon by a Wigner-Kirkwood expansion to
order $\hbar^{2}$. They found a lowering of the triple-point temperature
of about 1.4 K and of the critical temperature of about 1.8 K due
to quantum corrections. The quantum effect at the liquid-gas critical
point was studied by Young by the Feynman-Hibbs effective potential
method.\citep{young80} An interesting prediction was that quantum
effects modify the critical pressure, volume, and temperature, but
no effect was predicted on the critical exponents. The neon liquid-vapor
interface at the experimental triple point was studied by PI Monte
Carlo (MC) simulations and compared to classical simulations, showing
that the quantum interface is approximately one quarter of interatomic
spacing wider than the classical counterpart.\citep{broughton89}
The liquid-vapor equilibrium of neon has been studied by Gibbs ensemble
MC calculations by both the PI formulation,\citep{wang97} and a Wigner-Kirkwood
expansion. \citep{venka04} Quantum effects in the liquid-vapor equilibrium
were shown to be important and more significant in the liquid state.
The critical temperature in the quantum simulation was found about
8\% lower than in the classical limit.\citep{wang97} This value results
to be larger than that of 4\% predicted by a Wigner-Kirkwood expansion
in Ref. {[}\onlinecite{hansen69}]. The solid-liquid coexistence
line of neon has been studied using a Wigner-Kirkwood effective potential
at pressures between $10^{2}$ and $4\times10^{3}$ MPa, showing that
quantum effects are negligible in this range of pressures.\citep{solca98}
A recent PI MC simulation has shown that at ambient pressure the melting
temperature of neon is lowered by about 6 \% (i.e. 1.5 K) when quantum
effects are included.\citep{ramirez08}

In the present work we study the phase diagram of natural neon for
temperatures in the range 17-50 K and pressures between $10^{-2}$
and $2\times10^{3}$ bar by calculation of the phase coexistence between
solid, liquid, and gas phases. Moreover the VPIE of solid and liquid
phases at triple-point conditions and the TPIE of $^{20}$Ne and $^{22}$Ne
are also calculated and compared to experimental data. All results
are obtained by PI MC simulations. Phase coexistence is obtained by
requiring equality of the Gibbs free energy, $G$, of the phases in
equilibrium. The calculation of $G$ is performed using the Adiabatic
Switching (AS)\citep{watanabe90} and Reversible Scaling (RS)\citep{koning99}
approaches that are based on algorithms where the Hamiltonian (or
a state variable as the pressure or inverse temperature) changes along
the simulation run. The capability of both AS and RS methods to calculate
free energies in the context of PI simulations has been recently analyzed
in the study of neon melting, \citep{ramirez08} and we will make
extensive use of these free energy techniques in the present investigation.
The influence of quantum effects in the phase diagram of neon will
be quantified by comparing the predicted phase diagram calculated
with and without atomic quantum effects included. The interatomic
interaction between Ne atoms will be described by a Lennard-Jones
pair potential, in analogy to our previous studies.\citep{herrero01,herrero05,ramirez05,ramirez08}
We recognize that a better option for the neon pair potential is given
by the fit to ab initio electronic structure 
calculations,\citep{venka04,vogt01,nasrabad04}
however we have chosen the simple LJ pair potential for several reasons:
the phase diagram in the classical case can be checked against a large
number of previous LJ simulations,\citep{mastny07,barroso02,agrawal95}
quantum effects in the three coexistence lines (solid-gas, solid-liquid,
and liquid-gas) has not yet been calculated by PI simulations using
the same LJ potential parameters, and the isotope effects in the phase
diagram depend more on the kinetic energy than on the potential energy
parameters.

The structure of this paper is as follows. In Sec. \ref{sec:methodology}
we present the computational conditions employed in the PI simulations
and the technique used to evaluate the free energy as a function of
the isotope mass. The results obtained by PI and classical simulations
for the three coexistence lines of neon are presented in 
Sec.~\ref{sec:phasediagram}.
The phase diagram of neon is displayed both in the pressure-volume
($P-V$) and in the density-temperature ($\rho-T$) domains. The TPIE
of $^{20}$Ne and $^{22}$Ne is calculated and compared to experimental
data in Sec. \ref{sec:TPIE}, while the VPIE of both solid and liquid
phases of neon are derived in Sec. \ref{sec:VPIE}, including also
the comparison to experiment. Finally, we summarize our conclusions
in Sec. \ref{sec:conclusions}.

\section{methodology\label{sec:methodology}}

\subsection{Computational conditions\label{subsec:pi}}

In the PI formulation of statistical mechanics the partition function
is calculated through a discretization of the integral representing
the density matrix. This discretization serves to define cyclic paths
composed by a finite number $L$ of steps, that in the numerical simulation
translates into the appearance of $L$ replicas (or beads) of each
quantum particle. Then, the implementation of PI simulations relies
on an isomorphism between the quantum system and a classical one,
derived from the former by replacing each quantum particle (here,
atomic nucleus) by a ring polymer of $L$ classical particles, connected
by harmonic springs with a temperature- and mass-dependent force constant.
Details on this computational method are given 
elsewhere.\citep{feynman72,gillan88,ceperley95,chakravarty97}
In this work the technical implementation of the PI simulations in
the $NPT$ and $NVT$ ensembles ($N$ being the number of particles)
is identical to that presented in Ref. {[}\onlinecite{ramirez08}].

For the PI simulation of the phase diagram of natural neon, the rare
gas atoms were treated as particles of mass $m_{0}=20.18$ amu interacting
through a LJ potential with parameters $\epsilon=3.2135$ meV and
$\sigma=2.782$ \AA. The value of the parameter $\epsilon$ differs
from that employed in our study of the melting transition of neon
at ambient pressure,\citep{ramirez08} where we found that the classical
melting temperature ($T_{m,cl}=$24.95 K) was closer to the experimental
data ($T_{m}=$24.55 K) than the quantum result ($T_{m,q}=$23.43
K). Note that the classical melting temperature, $T_{m,cl}$, is proportional
to the value of $\epsilon$, and a significant deviation of $T_{m,q}$
from this proportionality is not expected in the quantum case. Thus,
the parameter $\epsilon$ used in Ref. {[}\onlinecite{ramirez08}]
has been increased by a scaling factor of $\alpha=$1.042, which is
roughly the factor needed to match experimental and calculated $T_{m,q}$
temperatures. The LJ potential was truncated for distances larger
than $r_{c}=$$4\sigma$ and shifted to zero to perform the MC sampling.
The internal energy was corrected by adding back this potential shift
in order to approximate the properties of the full LJ potential. 

PI MC simulations were performed on cubic cells containing 500 atoms,
assuming periodic boundary conditions. For the liquid, standard long-range
corrections were computed assuming that the pair correlation function
is unity, $g(r)=1$ for $r>r_{c}$, leading to well-known corrections
for the pressure and internal energies.\citep{johnson93} In the solid,
 the  long range correction was a static-lattice approximation (called
LR-s in Ref. {[}\onlinecite{ramirez08}]), that implies that atoms
beyond a certain number of neighbor shells are arranged as in the
perfect crystal ordering. From the value of the cutoff radius, we
determine that the first twelve neighbor shells (representing 248
neighbors of a central atom) were dynamically included in the simulation.
This approximation leads also to simple long-range corrections for
the pressure and internal energies.\citep{muser95,cuccoli97} The
virial estimator was used for the calculation of the kinetic 
energy.\citep{herman82}
Solid phase simulations were started with the atoms on their ideal
face-centered-cubic (fcc) positions, while the liquid and gas phase
simulations were initialized by using random positions. The number
of beads was set as $L=15$, which has been shown to be adequate for
the temperatures and pressures employed in this 
study.\citep{herrero01,ramirez05,herrero05}
Typical runs consisted of $3\times10^{4}$ MC steps (MCS) for equilibration,
followed by at least $10^{5}$ MCS for the calculation of equilibrium
properties. Some simulations were performed with larger runs up to
$6\times10^{5}$ MCS, to check the convergence of the results. Classical
results were obtained by setting the number of beads $L$=1 in our
PI MC codes.

\subsection{Free energy calculations\label{subsec:freeenergy}}

The thermodynamic integration (TI) is an standard technique that allows
us to obtain the free energy of a system by the calculation of the
reversible work needed to change the original system into a reference
state of known free energy.\citep{kirkwood35,allen,frenkel} The AS
method is an alternative to the standard TI, where the reversible
work is obtained by slowly changing the system Hamiltonian along the
simulation run.\citep{watanabe90} RS is an efficient technique to
obtain the free energy as a function of a state variable, typically
$T$ or $P$.\citep{koning99} The reversible work is calculated here,
in  way similar to the AS method, by slowly changing the state variable
along the simulation run. The PI implementation of both AS and RS
methods has been recently described in the study of solid-liquid coexistence
of neon at ambient pressure and will not be repeated here.\citep{ramirez08}
An additional free energy method employed in this work is the Morales
and Singer (MS) approach, that allows us to obtain the free energy
difference between a quantum system and its corresponding classical
limit in either the $NVT$ or $NPT$ ensembles.\citep{morales91,ramirez08}
Here we merely summarize how these methods fit into our calculation
procedure of the phase coexistence of neon.

The first step is to obtain the free energy of each phase (solid,
liquid, or gas) at some appropriate reference point in the ($T,P$)
plane.  An $NPT$ simulation of each phase at the chosen reference
point is run to determine its equilibrium volume. In the case of the
solid, the Helmholtz free energy, $F$, is then calculated by an AS
simulation in the $NVT$ ensemble, using the Einstein crystal as a
reference. For the quantum case the reference is a quantum Einstein
crystal,\citep{ramirez08} while in the classical limit a classical
Einstein crystal is used as reference.\citep{polson00}

In the case of a quantum fluid (either liquid or gas) , the free-energy
difference between the quantum and classical fluid is calculated with
the MS approach in either the $NVT$ or $NPT$ 
ensembles.\citep{morales91,ramirez08}
The excess free energy of the classical fluid  is then obtained from
the Johnson {\em et al.\/} parameterization.\citep{johnson93}
Finally, the ideal gas contribution is added to obtain the total free
energy of the quantum fluid.\citep{ramirez08}

In the case of the solid and liquid phases a finite size correction
was applied to the free energy $F_{N}$ obtained for $N=$500 atoms
by using the following relation
\begin{equation}
F=F_{N}-\alpha\frac{a(T)}{N}\;,\label{eq:FN}\end{equation}
where $F$ represents the infinite-size free energy, $a(T)$ is a
constant that depends linearly on temperature 
(see Ref. {[}\onlinecite{ramirez08}])
and the scaling factor $\alpha=1.042$ takes into account the change
in the $\epsilon$ parameter employed in this work and in 
Ref. {[}\onlinecite{ramirez08}].
The value of $a(T)$ for both solid and liquid phases of neon is obtained
from Tab. II and Eq. (21) in Ref. {[}\onlinecite{ramirez08}]. 

The coexistence temperature between two phases is then calculated
at a given pressure by a free energy approach, by which the free energy
of both phases is calculated over an overlapping range of temperatures
by the RS method. The coexistence temperature is then determined by
the point where the Gibbs free energy of both phases is identical.

\subsection{Free energy isotope effect\label{subsec:isotope_free_energy }}

The isotope effect in the phase coexistence of a substance is determined
by the free energy difference between isotopes. Given an isotope of
mass $m$, and assuming that the Helmholtz free energy of the natural
isotopic composition, $F(m_{0}),$ is known by any of the methods
summarized in Sec. \ref{subsec:freeenergy}, then the unknown free
energy, $F(m)$, can be calculated by the expression 
\begin{equation}
F(m)=F(m_{0})+\intop_{m_{0}}^{m}\frac{\partial F(m')}{\partial m'}dm'\;.\label{eq:Fm_exact}
\end{equation}
By considering the following thermodynamic definition of the kinetic
energy $K$, 
\begin{equation}
K(m')=-m'\frac{\partial F(m')}{\partial m'}\;,
\end{equation}
we get 
\begin{equation}
F(m)=F(m_{0})-\intop_{m_{0}}^{m}\frac{K(m')}{m'}dm'\;.\label{eq:Fm}
\end{equation}
This expression shows that the kinetic energy is the magnitude that
determines the free energy difference between two isotopes. The implementation
of the last equation in our PI simulations has been done by the AS
method.\citep{watanabe90} Thus, in a simulation run with a number
$M$ of MCS, the particle mass is changed uniformly from the initial
value $m_{0}$ to the final value $m$. The actual value of the mass
at the simulation step $i$ is, 
\begin{equation}
m_{i}=m_{0}+(i-1)\Delta m\;,
\end{equation}
where $\Delta m=(m-m_{0})/(M-1).$ The free energy difference, $\Delta F=F(m)-F(m_{o})$,
is then obtained as
\begin{equation}
\Delta F=-\sum_{{\displaystyle i=1}}^{M}w_{i}\frac{K_{v}(m_{i})}{m_{i}}\Delta m\;,
\end{equation}
where $K_{v}(m_{i})$ is the virial estimator of the kinetic energy
at step $i$. $w_{i}$ is an integration weight factor ($w_{i}=1$
except for $i$=1 or $i=N$, where $w_{i}=$0.5). A calculation of
isotope effects on solubilities of hydrogen has been done by a Fourier
PI relation equivalent to Eq. (\ref{eq:Fm}) by using TI. \citep{beck92}
In the case of the $NPT$ ensemble the mass dependence of the Gibbs
free energy can be derived from the following relation analogous to
Eq. (\ref{eq:Fm})
\begin{equation}
G(m)=G(m_{0})-\intop_{m_{0}}^{m}\frac{K(m')}{m'}dm'\;.\label{eq:Gm}
\end{equation}

For the calculation of VPIE at a given temperature it is useful to
determine the Gibbs free energy as a function of pressure, provided
that an initial value, $G(P_{0})$, is already known. We have employed
the following equation
\begin{equation}
G(P)=G(P_{0})+\intop_{P_{0}}^{P}VdP'\;,\label{eq:GP_exact}
\end{equation}
where we have made used of the relation $V=\partial G/\partial P$.
The calculation of the integral in the last equation is also done
by the RS method, i.e., the pressure $P'$ is uniformly changed between
$P_{0}$ and $P$ along the simulation run and the value of $G(P)$
is calculated {}``on the fly'' by solving numerically the integral
in Eq. (\ref{eq:GP_exact}) as the simulation proceeds.\citep{koning99}

\section{Phase diagram of neon\label{sec:phasediagram}}

The phase coexistence of natural neon has been derived from PI MC
simulations. We present the results obtained for the three coexistence
lines (solid-liquid, solid-gas, and liquid-gas). The studied temperature
range is between 17-50 K and pressures were varied between $10^{-2}$
and $2\times10^{3}$ bar. The quantum results are compared to the
classical limit and to available experimental data.

\subsection{Solid-liquid coexistence\label{subsec:sl}}

\begin{figure}
\vspace{-0.7cm}
\hspace{-0.5cm}
\includegraphics[height= 9cm]{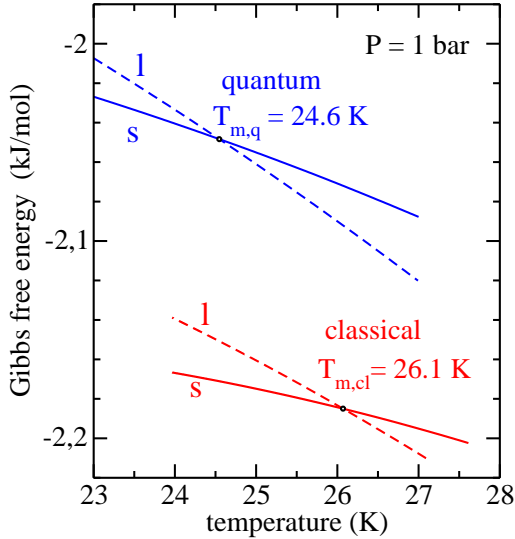}
\vspace{-1.5cm}
\caption{Melting temperature for
fcc neon at 1 bar obtained by the condition
of equal Gibbs free energy of the coexisting phases. Results are shown
for both quantum and classical descriptions of solid (s) and liquid
(l) phases. }
\label{fig:1}
\end{figure}

The free energy of both solid and liquid phases of neon over an overlapping
range of temperatures that encompasses the melting point is shown
in Fig.~\ref{fig:1} for $P$= 1 bar. The classical melting point
is found at 26.1 K. The main quantum effect in the free energy is
a shift of the classical data towards higher values. As this energy
shift results larger for the solid than for the liquid phase, the
melting temperature is displaced downward by about 1.5 K. The slopes
of the free energy curves are proportional to --$S$ ($S$ being the
entropy). From the figure, it is obvious that the entropy of the liquid
phase is larger than that of the solid. Less obvious to the eye is
that the entropy of both phases increases when quantum effects are
included. At melting conditions ($P$=1 bar, $T_{m,cl}=26.1$ K, and
$T_{m,q}=24.6$ K) the entropy of the solid phase amounts to 14.7
J/mol K (quantum) and 10.2 J/mol K (classical), while for the liquid
phase the corresponding values are 27.7 J/mol K (quantum) and 23.8
J/mol K (classical). Therefore the consideration of quantum effects
implies an increase of 44 \% in the solid entropy and of 16 \% in
the liquid one.

\begin{figure}
\vspace{-0.7cm}
\hspace{-0.5cm}
\includegraphics[height= 9cm]{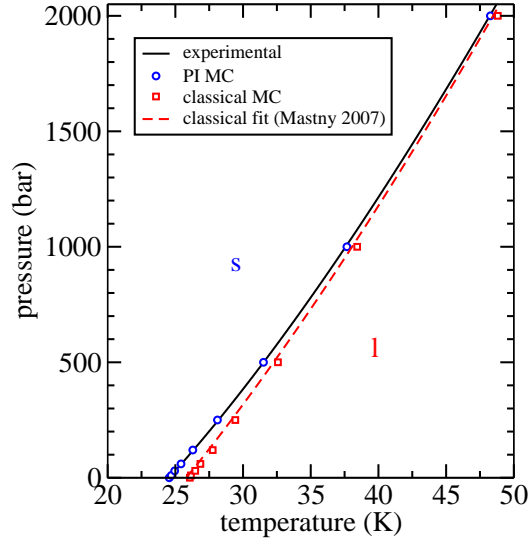}
\vspace{-1.5cm}
\caption{Solid-liquid (s-l)
coexistence line of neon in the $P-T$ domain.
The continuous line is the experimental result from Ref.
{[}\onlinecite{crawford71}
]. Open circles and squares are the results of our PI MC and classical
MC simulations, respectively. The broken line is the empirical fit
to classical simulations given in Ref. {[}\onlinecite{mastny07}]. }
\label{fig:2}
\end{figure}

The solid-liquid phase consistence has been calculated for 9 different
pressures up to 2 Kbar and the result for the melting line is shown
in Fig.~2. We observe that the quantum melting line is displaced with
respect to the classical limit. At zero pressure the quantum shift
in the melting temperature, $T_{m}$, amounts to 1.6 K (about 6 \%
of $T_{m}$), while at 2 Kbar this shift is only of about 0.6 K (about
1 \% of $T_{m}$). Therefore, the quantum effect in the melting temperature
decreases as pressure goes up. This fact is in line with the Wigner-Kirkwood
analysis of Ref. {[}\onlinecite{solca98}] for pressures between $10³$
and $2\times10^{4}$ bar. Our PI MC data show a reasonable overall
agreement to experimental results.\citep{crawford71} The agreement
is in part originated by our selection of the LJ parameter $\epsilon$
to fit the experimental melting temperature (24.55 K) at ambient pressure
(see Subsec. \ref{subsec:pi}). The broken line in the figure shows
the classical simulation results of Ref. {[}\onlinecite{mastny07}]
for the LJ parameters employed here. These data are slightly shifted
with respect to our classical melting line. We obtain somewhat higher
melting temperatures for a given pressure. A possible reason for this
discrepancy is that in both simulations the long range correction
used for the solid phase and the LJ potential cutoff are different.

\subsection{Solid-gas coexistence\label{subsec:sg}}

\begin{figure}
\vspace{-0.2cm}
\hspace{-0.5cm}
\includegraphics[height= 9cm]{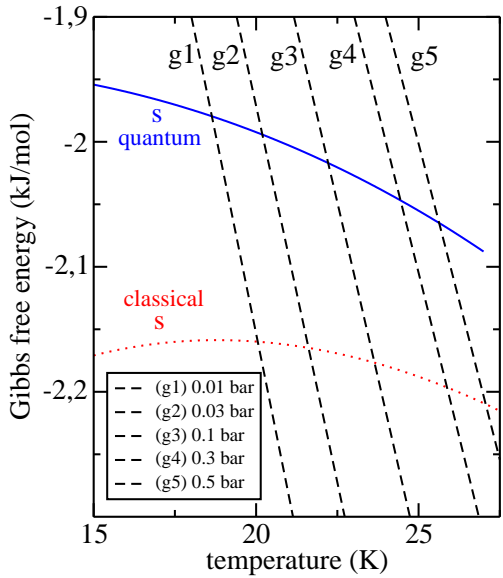}
\vspace{-1.5cm}
\caption{Gibbs free energies of
the solid (s) and gas (g) phases of neon as
a function of temperature. Gas results are shown for five different
pressures. The free energy of the solid corresponds to a pressure
of 1 bar. For the solid, the value of $G$ for pressures between 0
and 1 bar are indistinguishable at the scale of the figure. Results
 for both quantum and classical simulations of solid fcc neon are
shown. }
\label{fig:3}
\end{figure}

The Gibbs free energy $G$ of both solid and gas phases of neon is
shown as a function of temperature in Fig.~\ref{fig:3}. The results
derived for the gas from the Johnson parameterization are shown for
five pressures between $10^{-2}$ and $0.5$ bar. Neither quantum
effects nor deviations from ideal behavior are significant in the
displayed range of pressures for the gas free energies. The free energy
of the solid corresponds to the data calculated at a pressure of 1
bar. At the scale of this figure, solid free energies at lower pressures
are indistinguishable from the $P=1$ bar data and  are not represented
 (the $PV$ term of the Gibbs free energy is of the order of 1 J/mol
at these pressures). However, lower pressure results were calculated
for the solid phase to determine the solid-gas coexistence line. In
contrast to the gas behavior, quantum effects are important in the
solid, and thus both classical and quantum free energy results are
plotted in Fig.~\ref{fig:3}. It is interesting to note that at temperatures
about 19 K the classical free energy of the solid displays a maximum,
therefore below this temperature the entropy of the classical solid
becomes negative and in fact it would tend to $-\infty$ as temperature
goes down to zero. A simple way to derive this result is to work out
analytically a harmonic model for the classical solid. This behavior
of the classical entropy is known to be incompatible with the Nerst
theorem that constitutes the third principle of thermodynamics.

\begin{figure}
\vspace{-0.7cm}
\hspace{-0.5cm}
\includegraphics[height= 9cm]{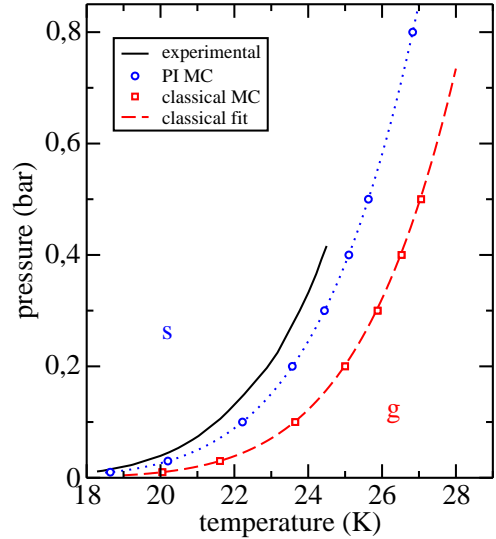}
\vspace{-1.5cm}
\caption{Solid-gas (s-g)
coexistence curve of neon in the $P-T$ domain. The
continuous line is the experimental result from Ref.
{[}\onlinecite{bigeleisen61}].
Open circles and squares are the results of our PI MC and classical
MC simulations, respectively. The dotted line is a polynomial fit
of fourth degree to the PI MC results. The broken line is a classical
fit obtained by an empirical EoS in Ref.
{[}\onlinecite{hoef02}].
\label{fig:4}}
\end{figure}

The crossing points of the solid and gas free energy curves, $G(T),$
at a given value of $P$ determine the sublimation curve of the solid.
We have calculated the solid-gas coexistence line at eighth values
of $P$ in the range between $10^{-2}$ and 0.8 bar. The results are
shown in Fig.~\ref{fig:4} and compared to available experimental
data up to the triple-point temperature of neon.\citep{bigeleisen61}
The PI MC results (open circles) display a constant shift of about
0.6 K with respect to the experimental data. We note that by the same
scaling considerations as presented in Subsec. \ref{subsec:pi} for
the LJ parameter $\epsilon$, a decrease in the value of  $\epsilon$
would produce a shift of our calculated solid-gas coexistence line
toward lower temperatures. In this way we might improve the agreement
of our sublimation line to experiment. However, the calculated melting
line would then appear also shifted toward lower temperatures, and
thus the good agreement to experimental data shown in Fig.~\ref{fig:2}
would be lost. We conclude that a simple LJ pair potential seems to
be unable to describe both solid-gas and solid-liquid coexistence
lines of neon in good agreement to experiment. 

Our classical result for the sublimation line of neon is also given
in Fig.~\ref{fig:4}. The shift with respect to the PI MC results
is entirely due to quantum effects of the solid phase, as the gas
behaves as an ideal gas for the displayed values of pressure and temperature.
The broken line in the figure shows the fit of van der Hoef of the
solid-gas coexistence using an empirical equation of state (EoS) for
the LJ solid and the Johnson parameterization for the gas. \citep{hoef02}
Our classical results are consistent with this fit.

\subsection{Liquid-gas coexistence\label{subsec:lg}}

\begin{figure}
\vspace{-0.5cm}
\hspace{-0.5cm}
\includegraphics[height= 9cm]{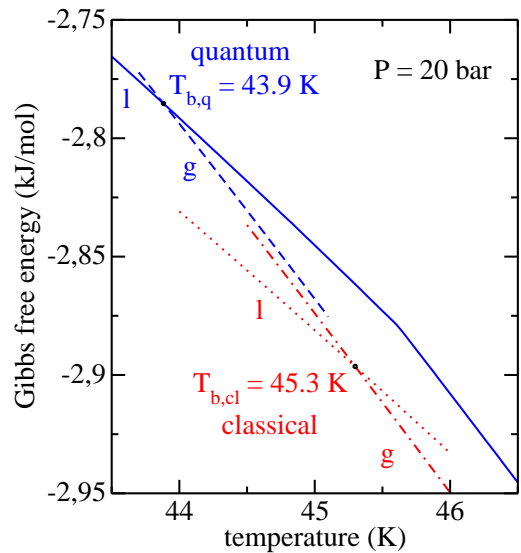}
\vspace{-1.0cm}
\caption{Determination of the
boiling point, $T_{b}$, of neon at 20 bar. The
calculated $G(T)$ curves for liquid (l) and gas (g) phases are
displayed
as a function of temperature. Results are displayed for both quantum
and classical simulations.
\label{fig:5}}
\end{figure}

The Gibbs free energy curves, $G(T)$, of liquid and gas neon phases
at pressure of 20 bar are displayed in Fig.~\ref{fig:5}. At this
pressure quantum corrections to the free energy of the gas are significant,
although smaller than in the case of the liquid phase. The shift in
the boiling temperature due to quantum effects amounts to 1.4 K. The
classical $G(T)$ curve for the gas phase was derived  from the Johnson
parameterization,\citep{johnson93} while for the liquid phase the
classical curve was obtained by the RS method started at a temperature
of $T=44$ K. We have proceeded in this way to check  the consistency
of our classical MC liquid simulations against the Johnson 
EoS.\citep{johnson93}

\begin{figure}
\vspace{-0.7cm}
\hspace{-0.5cm}
\includegraphics[height= 9cm]{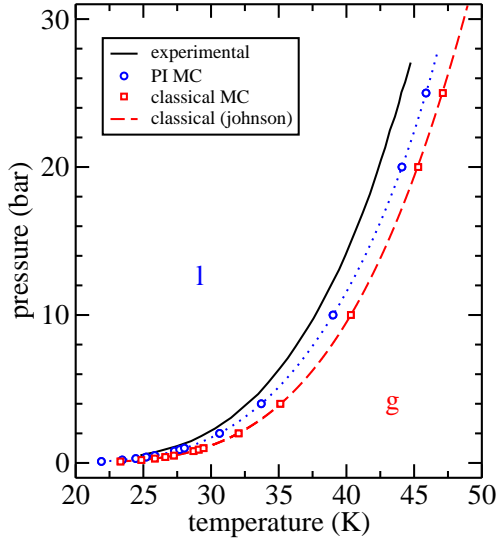}
\vspace{-1.5cm}
\caption{Liquid-gas (l-g)
coexistence curve of natural neon in the $P-T$ domain.
The continuous line is the experimental result from Ref.
{[}\onlinecite{rabinovich88}].
Open circles and squares are the results of our PI MC and classical
MC simulations, respectively. The dotted line is a polynomial fit
of sixth degree to the PI MC results. The broken line is the
coexistence
line obtained by using the parameterized EoS of
Johnson.\citep{johnson93}
\label{fig:6}}
\end{figure}

The liquid-gas coexistence line of natural neon is displayed in 
Fig.~\ref{fig:6}. Our classical results, derived from RS simulations of
the free energy in the liquid phase, are consistent with the coexistence
line directly obtained from the Johnson EoS.\citep{johnson93} The
PI MC boiling line is shifted with respect to the classical result.
Quantum corrections in the gas phase were found to be significant
at pressures above 5 bar. Our quantum results deviate from the experimental
boiling line.\citep{rabinovich88} Similar differences have been reported
in Gibbs ensemble MC simulations of the vapor-liquid phase coexistence
of neon, \citep{wang97} and the main reason for this discrepancy
seems to be related to the necessity of including three body terms
in the interaction potential.\citep{venka04,nasrabad04,vogt01}

\begin{figure}
\vspace{-1.6cm}
\hspace{-0.5cm}
\includegraphics[height= 9cm]{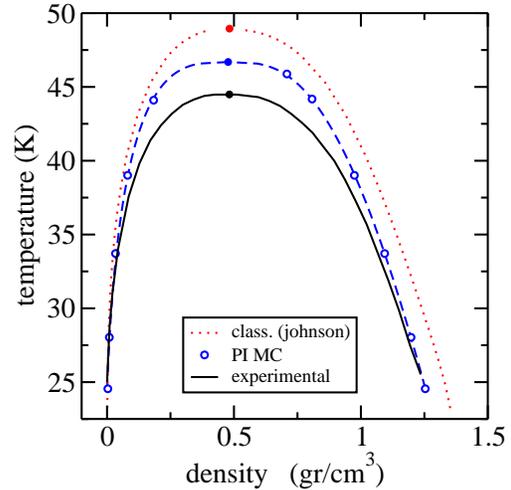}
\vspace{-1.4cm}
\caption{Density of coexistence of
vapor and liquid phases. Experimental data
are taken from Ref. {[}\onlinecite{rabinovich88} ], and classical
results from Ref. {[}\onlinecite{johnson93} ]. The broken line is
a fit of our PI MC results (open circles) to the scaling law presented
in the text {[}Eqs. (\ref{eq:rho1}) and (\ref{eq:rho2})]. Closed
circles correspond to the critical point in the different
approaches.
\label{fig:7}}
\end{figure}

\begin{table}
\caption{Comparison of the calculated critical point of neon with
experimental data.}
\begin{centering}
\label{tab:1}
\begin{tabular}{cccc}
\hline
  & experiment%
\footnote{From Ref {[}\onlinecite{rabinovich88} ].%
} & PI MC & classical%
\footnote{Johnson EoS.\citep{johnson93}%
}\tabularnewline
\hline
$T_{c}$(K) & 44.4 & 46.7 & 49.0\tabularnewline
$P_{c}$(bar) & 26.5 & 27.7 & 30.9\tabularnewline
$\rho_{c}$(g/cm$^{3}$) & 0.483 & 0.477 & 0.483\tabularnewline
\hline
\end{tabular}
\par\end{centering}
\end{table}

We have performed $NPT$ simulations of both liquid and vapor phases
of neon at coexistence conditions to obtain the saturation densities.
The critical temperature, $T_{c}$, has been then estimated by fitting
the calculated densities of the liquid and vapor phases, $\rho_{l}$
and $\rho_{g}$, to the scaling law\citep{frenkel}
\begin{equation}
\rho_{l}-\rho_{g}=b(T_{c}-T)^{\beta}\;,\label{eq:rho1}
\end{equation}
where $\beta=0.32$ is the non-classical critical exponent, which
is expected to be valid also in the quantum case,\citep{young80}
and $b$ is a constant. The values of $T_{c}$ and $b$ where obtained
by a least-squares fit of our PI MC results. The critical density,
$\rho_{c}$, was estimated by a fit to the law of rectilinear diameters,\citep{frenkel}
\begin{equation}
\frac{1}{2}(\rho_{l}+\rho_{g})=\rho_{c}+A(T-T_{c})\;.\label{eq:rho2}
\end{equation}
where the constant $A$ and the density, $\rho_{c}$, are the fitted
parameters. The PI MC estimation of the critical pressure, $P_{c},$
was obtained from the fit of the boiling line in Fig.~\ref{fig:6}
at temperature $T_{c}$. The critical parameters estimated from our
quantum simulations are compared to the experimental data and to the
Johnson EoS in Tab. \ref{tab:1}, and the results for the saturation
densities are presented in Fig.~\ref{fig:7}. The quantum correction
to the saturation density is of opposite sign for vapor and liquid
phases, while the vapor coexistence density increases, the liquid
one decreases. Although the PI MC saturation densities are closer
to experiment than the classical result, the deviation from experiment
is notorious, especially as one approaches the critical point. This
discrepancy is also found in previous simulations of the vapor-liquid
coexistence of neon using different pair potentials and quantum corrections,
and reflects the importance of three body terms in the interaction
potential to correctly describe phase coexistence near critical point
conditions. \citep{wang97,venka04,nasrabad04} From the difference
between calculated and experimental values of $T_{c}$ in Tab. \ref{tab:1}
one can infer that near critical point conditions both quantum corrections
and three body terms should be of similar importance.

\subsection{P-T domain\label{subsec:pv}}

\begin{figure}
\vspace{-0.7cm}
\hspace{-0.5cm}
\includegraphics[height= 9cm]{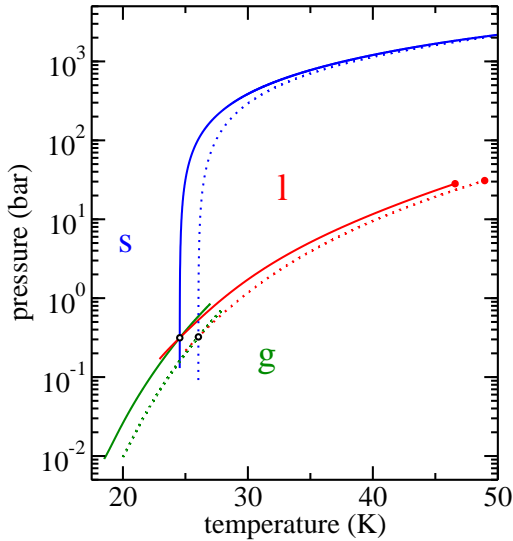}
\vspace{-1.5cm}
\caption{Phase diagram of natural
neon in the pressure-temperature domain.
Results are shown for the three coexistence lines between solid (s),
liquid (l), and gas (g) phases. Continuous lines are derived from
the PI MC results shown in Figs.~\ref{fig:2}, \ref{fig:4}, and
\ref{fig:6},
while dotted lines are derived from the corresponding classical data
in the same figures. Open circles represent the triple point and closed
ones the critical point.\label{fig:8}}
\end{figure}

\begin{table}
\caption{Comparison of predicted triple-point location of natural neon
with experimental data.}
\label{tab:2}
\begin{tabular}{cccc}
\hline
 & experiment & PI MC & classical\tabularnewline
\hline
$T_{tp}$ (K) & 24.553%
\footnote{Ref. {[}\onlinecite{furokawa72}].%
} & 24.55 & 26.04\tabularnewline
$P_{tp}$(bar)  & 0.433$^{a}$ & 0.315 & 0.323\tabularnewline
$\rho_{tp,s}$ (g/cm$^{3}$) & 1.424%
\footnote{Ref. {[}\onlinecite{vos91}].%
} & 1.429 & 1.494\tabularnewline
$\rho_{tp,l}$(g/cm$^{3}$) & 1.247$^{b}$ & 1.254 & 1.313\tabularnewline
$\rho_{tp,g}$(g/cm$^{3}$) & - & 0.0032 & 0.0031\tabularnewline
\hline
\end{tabular}
\end{table}

\begin{table}
\caption{PI MC results for thermodynamic properties of natural neon at
triple-point
conditions.}
\label{tab:3}
\begin{tabular}{ccccc}
\hline
 & $G$ (kJ/mol) & $PV$ (kJ/mol) & $U$ (kJ/mol) & $S$ (J/mol
K)\tabularnewline
\hline
solid  & -2.0492 & 0.0004 & -1.6883 & 14.68\tabularnewline
liquid  & -2.0492 & 0.0005 & -1.3719 & 27.57\tabularnewline
gas  & -2.0492 & 0.2002 & 0.2989 & 103.80\tabularnewline
\hline
\end{tabular}
\end{table}

The three calculated coexistence lines of natural neon are presented
in Fig.~\ref{fig:8} in the $P-T$ domain. Note that the pressure
is plotted using a logarithmic scale. Continuous lines were derived
from PI MC simulations, while dotted lines are classical results.
The triple point is determined by the crossing of the three coexistence
lines. For both quantum and classical results we find that the three
coexistence lines meet within a small temperature interval of 0.01
K and within a pressure interval of 0.002 bar. This small dispersion
in the triple point determination is an internal check of the consistency
of our results. Calculated triple-point properties are summarized
in Tab. \ref{tab:2}. The value of $T_{tp}$ predicted by the PI
MC results is in good agreement to experiment, while classical simulations
result in a higher value of $T_{tp}$ by about 1.5 K.  The equilibrium
densities of the solid and liquid phases calculated by PI MC are also
in good agreement to experiment. The calculated triple-point pressure,
$P_{tp},$ is lower than the experimental data, and we do not observe
any appreciable quantum correction to the value of $P_{tp}$. The
calculated values of $G$, $PV,$ internal energy ($U)$, and $S$
at triple-point conditions are summarized in Tab. \ref{tab:3}.

\subsection{T-$\rho$ domain \label{subsec:t_rho}.}

\begin{figure}
\vspace{-0.7cm}
\hspace{-0.5cm}
\includegraphics[height= 9cm]{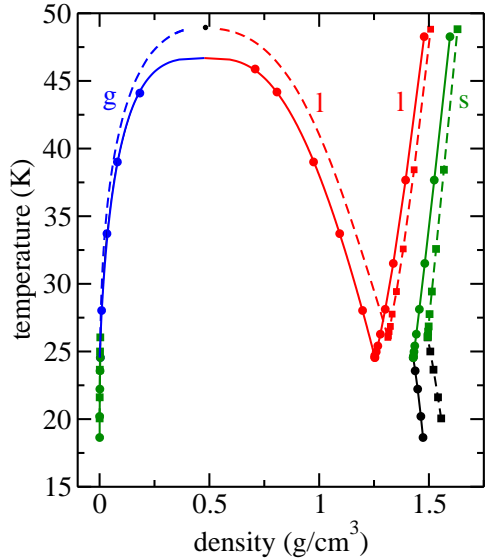}
\vspace{-1.4cm}
\caption{Phase diagram of natural neon in the temperature-density
domain. Closed
squares and circles are results obtained from classical MC and quantum
PI MC simulations, respectively. Classical vapor and liquid densities
were derived from the Johnson EoS,\citep{johnson93} while the
continuous
line through the quantum vapor and liquid densities is the fit to
the scaling law presented in the text {[}Eqs. (\ref{eq:rho1}) and
(\ref{eq:rho2})]. Results are shown for the three coexistence lines
between solid (s), liquid (l), and gas (g) phases.
\label{fig:9}}
\end{figure}

$NPT$ simulations of coexisting phases have been performed along
the three coexistence lines determined in the previous Subsections
to obtain the phase diagram of natural neon in the temperature-density
domain. The PI MC and classical results are presented in Fig.~\ref{fig:9}.
The largest deviation between classical and quantum results is found
for the solid phase, especially at temperatures below the triple point.
We observe that quantum corrections to the liquid coexistence density
at temperatures between the triple ($T_{tp})$ and critical points
($T_{c})$ results to be larger for the liquid-gas line than for the
liquid-solid one. This behavior is probably related to the trend that
quantum effects are comparatively less important as pressure increases.\citep{solca98}
The liquid at temperatures between $T_{tp}$ and $T_{c}$ is under
higher pressure if it is in equilibrium with the solid than if it
is in equilibrium with its vapor (see the $P-T$ phase diagram in
Fig.~\ref{fig:8} and note that the coexistence pressure increases
in both cases due to quantum effects).

\section{Triple-point isotope effect\label{sec:TPIE}}

We turn now to the study of the TPIE. We have previously obtained
that the triple-point pressure, $P_{tp}$, of neon is nearly the same
in both quantum and classical limits. Therefore one can reasonably
assume that $P_{tp}$ is nearly independent of the isotope mass. This
fact simplifies the calculation of the TPIE that will be derived from
a phase coexistence calculation at fixed pressure ($P_{tp}=0.315$
bar). We have arbitrarily chosen to calculate the coexistence between
solid and liquid phases, although any other two phases would be equally
appropriate. The melting point of the neon isotopes will be determined
by the same procedure shown in Fig.~\ref{fig:1} for natural neon. 

\begin{figure}
\vspace{-0.7cm}
\hspace{-0.5cm}
\includegraphics[height= 9cm]{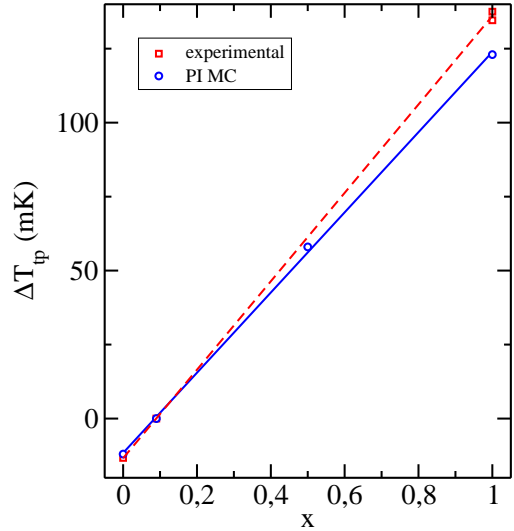}
\vspace{-1.5cm}
\caption{Triple-point temperature
differences for the different neon isotopes
as a function of the variable $x$ defined in the text {[}Eq.
(\ref{eq:x})].
PI MC results are compared to experimental data.
\label{fig:10}}
\end{figure}

\begin{table}
\caption{Free energy differences ($\Delta G_{m}$, in J/mol) between
neon isotopes
and natural neon calculated at the triple point of natural neon. The
calculated and experimental triple-point temperatures for each isotope
are collected in the last lines.}
\label{tab:4}
\begin{tabular}{cccc}
\hline
 & $^{20}$Ne & $^{21}$Ne & $^{22}$Ne\tabularnewline
\hline
$\Delta G_{m}$ (s) & 3.8 & -16.9 & -36.4\tabularnewline
$\Delta G_{m}$ (l)  & 3.6 & -16.1 & -34.8\tabularnewline
$T_{tp}$(K) (PI MC)  & 24.538 & 24.608 & 24.673\tabularnewline
$T_{tp}$(K) (exper.)%
\footnote{Ref. {[} \onlinecite{furokawa72}].%
}  & 24.540 & - & 24.687\tabularnewline
\hline
\end{tabular}
\end{table}

We take $(T_{tp},P_{tp})$ of natural neon as reference point to calculate
the Gibbs free energy of the solid and liquid phases of the neon isotope
with mass $m.$ The Gibbs free energy difference,
\begin{equation}
\Delta G_{m}=G(m)-G(m_{0})\;,
\end{equation}
between the isotope of mass $m$ and natural neon ($m_{0}$) is derived
by an AS calculation using Eq. (\ref{eq:Gm}). The free energy $G(m_{0})$
was given in Tab. \ref{tab:3}, and the results of $\Delta G_{m}$
for both solid and liquid phases are summarized in Tab. \ref{tab:4}
as a function of the isotope mass $m$. The triple-point temperatures,
$T_{tp}$, obtained as the melting point of the isotopes at $P_{tp}=0.315$
bar are also given in Tab. \ref{tab:4}. The difference between the
experimental triple-point temperatures of $^{22}$Ne and $^{20}$Ne
is 0.147 K, \citep{furokawa72} while our PI MC simulations predict
a value of 0.135 K. It is customary to represent the isotopic composition
of a neon sample of mass $m$ by a variable $x$, defined as
\begin{equation}
x=\frac{m-m_{20}}{2}\;,\label{eq:x}
\end{equation}
where $m_{20}=$20 amu. The PI MC results for the triple-point temperature
difference with respect to natural neon are presented in Fig.~\ref{fig:10}
as a function of $x.$ The calculated TPIE is compared to the experimental
data taken from Ref. {[}\onlinecite{pavese08}]. The linear behavior
experimentally found as a function of $x$ is reasonably reproduced
by our simulations, however our slope is smaller that the experimental
one.

\section{Vapor pressure isotope effect\label{sec:VPIE}}

\begin{figure}
\vspace{-1.5cm}
\hspace{-0.5cm}
\includegraphics[height= 9cm]{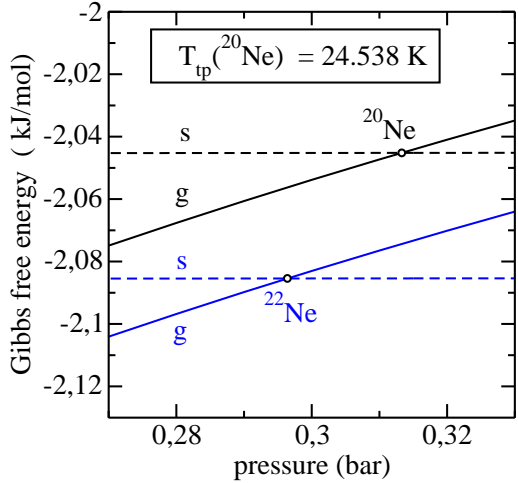}
\vspace{-1.5cm}
\caption{Determination of the vapor pressure of $^{20}$Ne and $^{22}$Ne
at the triple point of $^{20}$Ne by the study of the solid-gas
coexistence. }
\label{fig:11}
\end{figure}

\begin{table}
\caption{VPIE of $^{20}$Ne and $^{22}$Ne for solid (s) and liquid (l)
phases at triple-point conditions. $p'$and $p$ denote the vapor pressure
of $^{20}$Ne and $^{22}$Ne, respectively.}
\label{tab:5}
\begin{tabular}{ccc}
\hline
 & $\ln(p'/p)_{s}$%
\footnote{Values obtained at $T_{tp}$($^{20}$Ne)%
} & $\ln(p'/p')_{l}$%
\footnote{Values obtained at $T_{tp}$($^{22}$Ne)%
}\tabularnewline
\hline
PI MC  & 0.056 & 0.047\tabularnewline
exper.%
\footnote{Ref. {[}\onlinecite{bigeleisen61}]%
} & 0.058 & 0.047\tabularnewline
\hline
\end{tabular}
\end{table}

The VPIE is experimentally tabulated as the logarithmic ratio of the
isotopic vapor pressure, $\ln(p'/p)$, where $p'$and $p$ are the
vapor pressures of isotopes $m$ and $m'$, with $m>m'$. We have
calculated the VPIE for $^{22}$Ne and $^{20}$Ne at the critical
points of both isotopes. First we consider the critical temperature
of the lightest isotope, $^{20}$Ne. Note that this temperature is
lower that the triple-point temperature of $^{22}$Ne. This implies
that at this temperature the vapor of $^{22}$Ne may coexist only
with its solid phase (as a visual help see the phase diagram in 
Fig.~\ref{fig:8}, where there appear two critical points at different
temperatures). Therefore we have calculated the vapor pressure of
$^{22}$Ne and $^{20}$Ne at $T_{tp}$($^{20}$Ne )=24.538 K by the
study of the solid-gas coexistence. In Fig.~\ref{fig:11} we have
plotted the pressure dependence of the Gibbs free energy for both
solid and gas phases  of $^{22}$Ne and $^{20}$Ne. The free energy
of the solid was calculated by a RS simulation using Eq. (\ref{eq:GP_exact}),
while the free energy of the gas was obtained from the classical EoS.\citep{johnson93}
The vapor pressures of $^{20}$Ne, $p$' = 0.313 bar, and of $^{22}$Ne,
$p=$0.296 bar, are read from the crossing points in the figure. The
resulting value of $\ln(p'/p)$ is compared to available experimental
data in Tab. \ref{tab:5}.\citep{bigeleisen61} 

\begin{figure}
\vspace{-1.5cm}
\hspace{-0.5cm}
\includegraphics[height= 9cm]{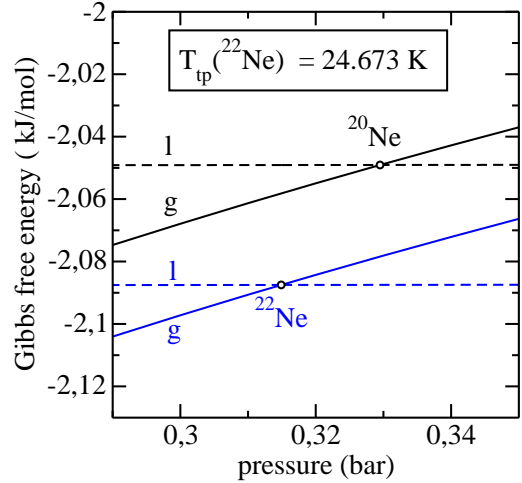}
\vspace{-1.5cm}
\caption{Determination of the
vapor pressure of $^{20}$Ne and $^{22}$Ne at
the triple point of $^{22}$Ne by the study of the liquid-gas
coexistence.}
\label{fig:12}
\end{figure}

The VPIE at the triple point temperature of the heavier isotope, $^{22}$Ne,
has been calculated by the study of the liquid-gas coexistence shown
in Fig.~\ref{fig:12}. Gibbs free energies for the liquid phases were
derived from PI MC simulations while the gas phase results correspond
to the Johnson parameterization. \citep{johnson93} Here the vapor
pressure of $^{20}$Ne and $^{22}$Ne are $p$' = 0.330 bar and $p=$0.315
bar, and the calculated value of $\ln(p'/p)$ shows an excellent agreement
to the experimental data (see Tab. \ref{tab:5}).\citep{bigeleisen61,tew08}
Our simulations correctly reproduce that the VPIE is larger in the
solid than in the liquid phase.

\section{Conclusions\label{sec:conclusions} }

The calculation of the phase diagram of neon by means of path integral
simulations demonstrates that free energy techniques, previously used
in classical simulations, can be also effectively employed to study
quantum problems. In particular, we have made extensive used of the
Adiabatic Switching\citep{watanabe90} and Reversible Scaling\citep{koning99}
methods, that are based on algorithms where either the Hamiltonian,
a state variable (pressure, temperature) or even an atomic mass are
adiabatically changed along a simulation run. As long as this change
is performed slowly (adiabatically), the associated reversible work
is equal to the free energy difference between initial and final states.
These free energy algorithms can be easily implemented in any code
prepared for equilibrium simulations.

Our quantum simulations of the phase diagram of neon have focused
on the study of two types of effects. The first one is a theoretical
quantity not accessible by experimental measurements. This {}``quantum
effect'' is defined as the difference in phase coexistence properties
obtained by applying either quantum or classical statistical mechanics.
Such effects are not experimentally accessible, as there is no way
to perform coexistence measurements of neon phases in the classical
limit. Therefore, the calculation of these effects is interesting
only if we want to quantify the errors of a classical treatment of
phase coexistence. This interest exists because historically the approaches
used to study phase coexistence of rare gas atoms were based on classical
statistical mechanics. The second studied effect is real, in the sense
that it can be experimentally measured. We refer here to isotope effects
in the phase coexistence of neon, such as the  triple-point temperature
or vapor pressure. Classical statistical mechanics incorrectly predicts
that phase coexistence is independent of the atomic mass, and therefore
an isotope effect is classically forbidden. 

We have found that quantum effects in the phase diagram of neon are
significant at pressures below $2\times10^{3}$ bar. The whole solid-gas
and liquid-gas coexistence lines are found  below this limit. The
main quantum effect found in the $P-V$ domain for these two lines
is a shift of about 1.5 K toward lower temperatures. For the case
of solid-liquid coexistence, a temperature shift is found that decreases
with pressure, from a value of 1.5 K at triple point conditions to
about 0.6 K at $2\times10^{3}$ bar. No significant quantum effects
are found in the value of the triple-point pressure of neon.

The employed Lennard-Jones parameters  have been able to reproduce
reasonably the experimental melting line of neon in the studied pressure
range up to $2\times10^{3}$ bar. However, the calculated solid-gas
line shows a rigid shift of about 0.6 K toward higher temperatures
when compared to experimental data. By decreasing the value of the
Lennard-Jones parameter $\epsilon$ it is possible to shift the solid-gas
line towards lower temperatures, but the solid-liquid line is shifted
also by the same amount. Therefore it seems impossible to reproduce
correctly both coexistence lines by employing a simple Lennard-Jones
pair potential. The consideration of three body terms in the interaction
potential is likely a necessary step to achieve better agreement to
experiment. This consideration applies also for the liquid-gas coexistence,
specially when approaching critical point conditions. The triple point
is reasonable reproduced by our quantum simulations, in particular
the triple-point temperature and coexisting densities, while the predicted
triple-point pressure is too low.

Our study of isotope effects in the triple-point temperature and vapor
pressure of neon shows good agreement to experimental data. These
isotope effects depend on the free energy difference between neon
isotopes, that is a function of the kinetic energy and therefore not
as dependent on the details of the interaction potential. The calculated
difference between the triple-point temperatures of $^{22}$Ne and
$^{20}$Ne was 0.135 K, while the experimental result is 
0.147 K.\citep{furokawa72}
Our quantum simulations correctly reproduce that $^{20}$Ne is more
volatile than $^{22}$Ne and the calculated vapor-pressure isotope
effect in both solid and liquid phases at triple point conditions
show near quantitative agreement to experimental data.

\acknowledgments This work was supported by Ministerio de Ciencia
e Innovaci\'on (Spain) through Grant No. FIS2006-12117-C03 and by CAM
through project S-0505/ESP/000237. The authors benefited from discussions
with E. R. Hern\'andez and A. Antonelli.

\end{document}